\documentclass[aps,pra,showpacs,superscriptaddress,preprint]{revtex4}
\usepackage{amssymb}
\usepackage{amsmath}
\usepackage{graphicx}

\catcode`ð=\active
 \defð{\u{g}}
 \catcode`Ð=\active
 \defÐ{\u{G}}
 \catcode`Ý=\active
\defÝ{\. I}
 \catcode`ö=\active
\defö{\"{o}}
 \catcode`Ö=\active
 \defÖ{\"O}
 \catcode`ü=\active
 \defü{\"{u}}
 \catcode`Ü=\active
 \defÜ{\"{U}}
 \catcode`Þ=\active
\defÞ{\c{S}}
 \catcode`þ=\active
 \defþ{\c{s}}
 \catcode`ý=\active
 \defý{{\i}}
 \catcode`ç=\active
\defç{\d{c}}
 \catcode`Ç=\active
\defÇ{\d{C}}

\begin{document}

\title{Solutions of the spatially-dependent mass Dirac equation with the
spin and pseudo-spin symmetry for the Coulomb-like potential }
\author{Sameer M. Ikhdair}
\email[E-mail: ]{sikhdair@neu.edu.tr}
\affiliation{Department of Physics, Near East University, Nicosia, North Cyprus, Turkey}
\author{Ramazan Sever}
\email[E-mail: ]{sever@metu.edu.tr}
\affiliation{Department of Physics, Middle East Technical University, 06531, Ankara,Turkey}
\date{%
\today%
}

\begin{abstract}
We study the effect of spatially dependent mass function over the solution
of the Dirac equation with the Coulomb potential in the ($3+1$)-dimensions
for any arbitrary spin-orbit $\kappa $ state$.$ In the framework of the spin
and pseudospin symmetry concept, the analytic bound state energy eigenvalues
and the corresponding upper and lower two-component spinors of the two Dirac
particles are obtained by means of the Nikiforov-Uvarov method, in closed
form. This physical choice of the mass function leads to an exact analytical
solution for the pseudospin part of the Dirac equation. The special cases $%
\kappa =\pm 1$ ($l=\widetilde{l}=0,$ i.e., $s$-wave)$,$ the constant mass
and the non-relativistic limits are briefly investigated.

Keywords: Dirac equation, spin symmetry, pseudospin symmetry, bound states,
Coulomb potential, spatially-dependent mass, Nikiforov-Uvarov method.
\end{abstract}

\pacs{03.65.Pm; 03.65.Ge; 02.30.Gp}
\maketitle

\newpage

\section{Introduction}

Within the framework of the Dirac equation the spin symmetry arises if the
magnitude of the attractive scalar potential $S(r)$ and repulsive vector
potential are nearly equal, $S(r)\sim V(r)$ in nuclei (\textit{i.e.}, when
the difference potential $\Delta (r)=V(r)-S(r)=A=$ constant$).$ However, the
pseudospin symmetry occurs when $S(r)\sim -V(r)$ are nearly equal (\textit{%
i.e.}, when the sum potential $\Sigma (r)=V(r)+S(r)=A=$ constant$)$ [1-3]$.$
The spin symmetry is relevant for mesons [4]. The pseudospin symmetry
concept has been applied to many systems in nuclear physics and related
areas [2-7]. It has also been used to explain features of deformed nuclei
[8], the super-deformation [9] and to establish an effective nuclear
shell-model scheme [5,6,10]. The pseudospin symmetry introduced in nuclear
theory refers to a quasi-degeneracy of the single-nucleon doublets and can
be characterized with the non-relativistic quantum numbers $(n,l,j=l+1/2)$
and $(n-1,l+2,j=l+3/2),$ where $n,$ $l$ and $j$ are the single-nucleon
radial, orbital and total angular momentum quantum numbers for a single
particle, respectively [5,6]. The total angular momentum is given as $j=%
\widetilde{l}+\widetilde{s},$ where $\widetilde{l}=l+1$ is a pseudo-angular
momentum and $\widetilde{s}=1/2$ is a pseudospin angular momentum. In real
nuclei, the pseudospin symmetry is only an approximation and the quality of
approximation depends on the pseudo-centrifugal potential and pseudospin
orbital potential [11]. In ref. \ [12], Alhaidari \textit{et al}.
investigated in detail the physical interpretation on the three-dimensional
Dirac equation in the presence of exact spin symmetry limitation $\Delta
(r)=0$ and pseudospin symmetry\ limitation $\Sigma (r)=0.$

Some authors have applied the pseudospin symmetry on several physical
potentials, such as the harmonic oscillator [12-16], the Woods-Saxon
potential [17], the Morse potential [18-20], the Hulth\'{e}n potential [21],
the Eckart potential [22-24], the molecular diatomic three-parameter
potential [25], the P\"{o}schl-Teller potential [26] and the Rosen-Morse
potential [27].

On the other hand, the problem of the spatially-dependent effective mass is
presenting a growing interest along the last few years [28-31]. Many authors
have used different methods to study the partially exactly solvable and
exactly solvable Schr\"{o}dinger, Klein-Gordon (KG) and Dirac equations in
the presence of variable mass having a suitable mass distribution functions
in $1D,$ $3D$ and/or any arbitrary $D$-dimensional cases for various type
potentials [32-45]. In the context of spatially-dependent mass, we have used
and applied a recently proposed approximation scheme [45] for the
centrifugal term to find a quasi-exact analytic bound-state solution of the
radial KG equation with spatially-dependent effective mass for scalar and
vector Hulth\'{e}n potentials in any arbitrary dimensional space $D$ and
orbital angular momentum quantum number $l$ within the framework of the
Nikiforov-Uvarov (NU) method [46,47]. In addition, the $s$-wave bound state
solution of the (1$+1$)-dimensional KG equation with mass inversely
proportional to the distance from the force center for the inversely linear
potential was obtained [48]. Two particular cases were studied, the exact
spin symmetry when the vector potential and the scalar potential are equal
in magnitude $S(r)=V(r)$ and the pseudospin symmetry when the vector
potential is equal to the scalar potential in magnitude but not in sign $%
S(r)=-V(r).$ Very recently, the study of ref. [48] has also been extended by
us [49] to study the bound-state solutions of the ($3+1$)-dimensional KG
equation with position-dependent bosonic mass function $m(r)=m_{0}\left(
1+\lambda _{0}br^{-1}\right) ,$ where $r\neq 0$ for the attractive scalar
Coulomb-like potential $S(r)=-\kappa _{s}r^{-1}$ with $\kappa _{s}=\hbar
cq_{s}$ being the coupling constant, taking into consideration the general
mixings of scalar and vector Lorentz structure potential.

In the present paper, our aim is to study the exact solution of the
spatially dependent Dirac equation with the Coulomb-like field potential for
any arbitrary spin-orbit quantum number $\kappa .$ Under the conditions of
the spin symmetry $S(r)\sim V(r)$ and pseudospin symmetry $S(r)\sim -V(r)$,
we investigate the bound state energy eigenvalues and corresponding upper
and lower spinor wave functions in the framework of the NU method. We also
show that the spin and pseudo-spin symmetry Dirac solutions can be reduced
to the $S(r)=V(r)$ and $S(r)=-V(r)$ Klein-Gordon solutions [49] in the cases
of exact spin symmetry limitation $\Delta (r)=0$ and pseudospin symmetry
limitation $\Sigma (r)=0,$ respectively. Further, the solutions of the Dirac
equation with constant mass can be also generated from the general solution
when the constant $b$ in the mass function is set to zero.

The paper is organized as follows. In sect. 2, we outline the NU method.
Section 3 is devoted for the analytic bound state solutions of the ($3+1$%
)-dimensional Dirac equation with spatially dependent mass function for the
quantum system obtained by means of the NU method. The spin symmetry and
pseudo-spin symmetry solutions are investigated. In sect. 3, we study the
cases $\kappa =\pm 1$ ($l=\widetilde{l}=0,$ $s$-wave), the constant mass and
the non-relativistic limit and compare with other wave equations and models.
Finally, the relevant conclusions are given in sect. 4.

\section{NU Method}

The NU method is briefly outlined here and details can be found in ref.
[46]. This method was proposed to solve the second-order differential
equation of hypergeometric-type:
\begin{equation}
\psi _{n}^{\prime \prime }(r)+\frac{\widetilde{\tau }(r)}{\sigma (r)}\psi
_{n}^{\prime }(r)+\frac{\widetilde{\sigma }(r)}{\sigma ^{2}(r)}\psi
_{n}(r)=0,
\end{equation}%
where $\sigma (r)$ and $\widetilde{\sigma }(r)$ are polynomials, at most, of
second-degree, and $\widetilde{\tau }(r)$ is a first-degree polynomial. In
order to find a particular solution for eq. (1), let us decompose the wave
function $\psi _{n}(r)$ as follows:%
\begin{equation}
\psi _{n}(r)=\phi (r)y_{n}(r),
\end{equation}%
and use%
\begin{equation}
\left[ \sigma (r)\rho (r)\right] ^{\prime }=\tau (r)\rho (r),
\end{equation}%
to reduce eq. (1) to the form%
\begin{equation}
\sigma (r)y_{n}^{\prime \prime }(r)+\tau (r)y_{n}^{\prime }(r)+\lambda
y_{n}(r)=0,
\end{equation}%
with%
\begin{equation}
\tau (r)=\widetilde{\tau }(r)+2\pi (r),\text{ }\tau ^{\prime }(r)<0,
\end{equation}%
where the prime denotes the differentiation with respect to $r.$ One is
looking for a family of solutions corresponding to%
\begin{equation}
\lambda =\lambda _{n}=-n\tau ^{\prime }(r)-\frac{1}{2}n\left( n-1\right)
\sigma ^{\prime \prime }(r),\ \ \ n=0,1,2,\cdots ,
\end{equation}%
The $y_{n}(r)$ can be expressed in terms of the Rodrigues relation:%
\begin{equation}
y_{n}(r)=\frac{B_{n}}{\rho (r)}\frac{d^{n}}{dr^{n}}\left[ \sigma ^{n}(r)\rho
(r)\right] ,
\end{equation}%
where $B_{n}$ is the normalization constant and the weight function $\rho
(r) $ is the solution of the differential equation (3). The other part of
the wave function (2) must satisfy the following logarithmic equation%
\begin{equation}
\frac{\phi ^{\prime }(r)}{\phi (r)}=\frac{\pi (r)}{\sigma (r)}.
\end{equation}%
By defining
\begin{equation}
k=\lambda -\pi ^{\prime }(r).
\end{equation}%
one obtains the polynomial

\begin{equation}
\pi (r)=\frac{1}{2}\left[ \sigma ^{\prime }(r)-\widetilde{\tau }(r)\right]
\pm \sqrt{\frac{1}{4}\left[ \sigma ^{\prime }(r)-\widetilde{\tau }(r)\right]
^{2}-\widetilde{\sigma }(r)+k\sigma (r)},
\end{equation}%
where $\pi (r)$ is a parameter at most of order $1.$ The expression under
the square root sign in the above equation can be arranged as a polynomial
of second order where its discriminant is zero. Hence, an equation for $k$
is being obtained. After solving such an equation, the $k$ values are
determined through the NU method. \

In this context, we may also derive a parametric generalization from the NU
method valid for most potentials under consideration. To do this, we begin
by writting the hypergeometric equation in general parametric form as
\begin{equation}
\left[ r\left( c_{3}-c_{4}r\right) \right] ^{2}\psi _{n}^{\prime \prime }(r)+%
\left[ r\left( c_{3}-c_{4}r\right) \left( c_{1}-c_{2}r\right) \right] \psi
_{n}^{\prime }(r)+\left( -\xi _{1}r^{2}+\xi _{2}r-\xi _{3}\right) \psi
_{n}(r)=0,
\end{equation}%
with%
\begin{equation}
\widetilde{\tau }(r)=c_{1}-c_{2}r,
\end{equation}%
\begin{equation}
\sigma (r)=r\left( c_{3}-c_{4}r\right) ,
\end{equation}%
\begin{equation}
\widetilde{\sigma }(r)=-\xi _{1}r^{2}+\xi _{2}r-\xi _{3},
\end{equation}%
where the coefficients $c_{i}$ ($i=1,2,3,4$) and the analytic expressions $%
\xi _{j}$ ($j=1,2,3$) are calculated for the potential model under
consideration. Overmore, comparing eq. (11) with it's counterpart eq. (1),
we obtain the analytic polynomials, energy equation and wave functions
together with the associated coefficients expressed in general parameteric
form in Appendix A.

\section{Analytic Solution of the Dirac-Coulomb-Like Problem}

In spherical coordinates, the spatially-dependent mass Dirac equation for
fermionic massive spin-$1/2$ particles interacting with arbitrary scalar
potential $S(r)$ and the time-component $V(r)$ of a four-vector potential
can be expressed as [27,50-53]
\begin{equation}
\left[ c\mathbf{\alpha }\cdot \mathbf{p+\beta }\left( m(r)c^{2}+S(r)\right)
+V(r)-E\right] \psi _{n\kappa }(\mathbf{r})=0,\text{ }\psi _{n\kappa }(%
\mathbf{r})=\psi _{n\kappa }(r,\theta ,\phi ),
\end{equation}%
where $E$ is the relativistic energy of the system, $m(r)$ is the
spatially-dependent mass of the fermionic particle, $\mathbf{p}=-i\mathbf{%
\nabla }$ is the momentum operator, and $\mathbf{\alpha }$ and $\mathbf{%
\beta }$ are $4\times 4$ Dirac matrices, which have the following forms,
respectively [50-53]%
\begin{equation}
\mathbf{\alpha =}\left(
\begin{array}{cc}
0 & \mathbf{\sigma } \\
\mathbf{\sigma } & 0%
\end{array}%
\right) ,\text{ }\mathbf{\beta =}\left(
\begin{array}{cc}
\mathbf{I} & 0 \\
0 & -\mathbf{I}%
\end{array}%
\right) ,\text{ }\mathbf{P}=-i\hbar \mathbf{\nabla },
\end{equation}%
where $\mathbf{I}$ denotes the $2\times 2$ identity matrix and $\mathbf{%
\sigma }$ are three-vector Pauli spin matrices%
\begin{equation}
\sigma _{1}\mathbf{=}\left(
\begin{array}{cc}
0 & 1 \\
1 & 0%
\end{array}%
\right) ,\text{ }\sigma _{2}\mathbf{=}\left(
\begin{array}{cc}
0 & -i \\
i & 0%
\end{array}%
\right) ,\text{ }\sigma _{3}\mathbf{=}\left(
\begin{array}{cc}
1 & 0 \\
0 & -1%
\end{array}%
\right) .
\end{equation}%
For a particle in a spherical (central) field, the total angular momentum
operator $\mathbf{J}$ and the spin-orbit matrix operator $\widehat{\mathbf{K}%
}=-\mathbf{\beta }\left( \mathbf{\sigma }\cdot \mathbf{L}+\mathbf{I}\right) $
commute with the Dirac Hamiltonian, where $\mathbf{L}$ is the orbital
angular momentum operator. For a given total angular momentum $j,$ the
eigenvalues of $\widehat{\mathbf{K}}$ are $\kappa =-(j+1/2)$ for aligned
spin ($s_{1/2},$ $p_{3/2},$ \textit{etc.}) and $\kappa =j+1/2$ for unaligned
spin ($p_{1/2},$ $d_{3/2},$ \textit{etc.}). The spinor wave functions can be
classified according to the radial quantum number $n$ and the spin-orbit
quantum number $\kappa $ and can be written using the Pauli-Dirac
representation:%
\begin{equation}
\psi _{n\kappa }(\mathbf{r})=\frac{1}{r}\left(
\begin{array}{c}
F_{n\kappa }(r)Y_{jm}^{l}(\theta ,\phi ) \\
iG_{n\kappa }(r)Y_{jm}^{\widetilde{l}}(\theta ,\phi )%
\end{array}%
\right) ,
\end{equation}%
where $F_{n\kappa }(r)$ and $G_{n\kappa }(r)$ are the radial wave functions
of the upper- and lower-spinor components, respectively, $Y_{jm}^{l}(\theta
,\phi )$ and $Y_{jm}^{\widetilde{l}}(\theta ,\phi )$ are the spherical
harmonic functions coupled to the total angular momentum $j$ and it's
projection $m$ on the $z$ axis. The orbital and pseudo-orbital angular
momentum quantum numbers for spin symmetry $l$ and pseudospin symmetry $%
\widetilde{l}$ refer to the upper- and lower-components, respectively. For a
given spin-orbit quantum number $\kappa =\pm 1,\pm 2,\cdots ,$ the orbital
angular momentum and pseudo-orbital angular momentum are given by $%
l=\left\vert \kappa +1/2\right\vert -1/2$ and $\widetilde{l}=\left\vert
\kappa -1/2\right\vert -1/2,$ respectively$.$ The quasi-degenerate doublet
structure can be expressed in terms of a pseudo-spin angular momentum $%
\widetilde{s}=1/2$ and pseudo-orbital angular momentum $\widetilde{l}$ which
is defined as $\widetilde{l}$ $=l+1$ for aligned spin $j=\widetilde{l}-1/2$
and $\widetilde{l}$ $=l-1$ for unaligned spin $\ j=\widetilde{l}+1/2.$ For
example, ($3s_{1/2},2d_{3/2}$) and ($3\widetilde{p}_{1/2},2\widetilde{p}%
_{3/2}$) can be considered as pseudospin doublets.

Substituting eq. (18) into eq. (15), we obtain two radial coupled Dirac
equations for the spinor components
\begin{subequations}
\begin{equation}
\left( \frac{d}{dr}+\frac{\kappa }{r}\right) F_{n\kappa }(r)=\left(
m(r)c^{2}+E_{n\kappa }-\Delta (r)\right) G_{n\kappa }(r),
\end{equation}%
\begin{equation}
\left( \frac{d}{dr}-\frac{\kappa }{r}\right) G_{n\kappa }(r)=\left(
m(r)c^{2}-E_{n\kappa }+\Sigma (r)\right) F_{n\kappa }(r),
\end{equation}%
where $\Delta (r)=V(r)-S(r)$ and $\Sigma (r)=V(r)+S(r)$ are the difference
and sum potentials, respectively. Eliminating $G_{n\kappa }(r)$ in eq. (19a)
and $F_{n\kappa }(r)$ into eq. (19b), we get two second-order differential
equations for the upper and lower spinor components as
\end{subequations}
\begin{subequations}
\begin{equation}
\left\{ \frac{d^{2}}{dr^{2}}-\frac{\kappa \left( \kappa +1\right) }{r^{2}}-%
\frac{1}{\hbar ^{2}c^{2}}\left[ U_{-}(r)U_{+}(r)+\frac{g_{-}(r)}{U_{-}(r)}%
\left( \frac{d}{dr}+\frac{\kappa }{r}\right) \right] \right\} F_{n\kappa
}(r)=0,
\end{equation}%
\begin{equation}
\left\{ \frac{d^{2}}{dr^{2}}-\frac{\kappa \left( \kappa -1\right) }{r^{2}}-%
\frac{1}{\hbar ^{2}c^{2}}\left[ U_{-}(r)U_{+}(r)+\frac{g_{+}(r)}{U_{+}(r)}%
\left( \frac{d}{dr}+\frac{\kappa }{r}\right) \right] \right\} G_{n\kappa
}(r)=0,
\end{equation}%
where $U_{-}(r)=m(r)c^{2}+E_{n\kappa }-\Delta (r)$ and $%
U_{+}(r)=m(r)c^{2}-E_{n\kappa }+\Sigma (r),$ are the difference and the sum
functions, respectively. Also, $g_{-}(r)=c^{2}\frac{dm(r)}{dr}-\frac{d\Delta
(r)}{dr}$ and$\ $ $g_{+}(r)=c^{2}\frac{dm(r)}{dr}+\frac{d\Sigma (r)}{dr}$
being the derivative of the mass function minus the difference potential and
the derivative of the mass function plus the sum potential, respectively.
From the above equations, the energy eigenvalues depend on the quantum
numbers $n$ and $\kappa $, and also the pseudo-orbital angular quantum
number $\widetilde{l}$ according to $\kappa (\kappa -1)=\widetilde{l}(%
\widetilde{l}+1),$ which implies that $j=\widetilde{l}\pm 1/2$ are
degenerate for $\widetilde{l}\neq 0.$

At this stage, we take the vector potential in the form of an attractive
Coulomb-like field as
\end{subequations}
\begin{equation}
\Sigma (r)=V(r)=-\frac{\hbar cq_{v}}{r},\text{ }q_{v}=q,\text{ }r\neq 0,
\end{equation}%
where $q_{v}$ is being a vector dimensionless real parameter coupling
constant and $\hbar c$ is being a constant with $J.fm$ dimension. Equations
(20a) and (20b) can not be solved analytically because of the last term in
the equations, we find it convenient to solve the mathematical relation $%
c^{2}dm(r)/dr=-d\Sigma (r)/dr=-dV(r)/dr$ for the sake of eliminating this
term. We find out that mass function should be taken as
\begin{subequations}
\begin{equation}
m(r)=m_{0}+m_{1}/r,
\end{equation}%
\begin{equation}
m_{1}=m_{0}\lambda _{0}b,\text{ }\lambda _{0}=\hbar /m_{0}c,
\end{equation}%
with $m_{0}$ and $m_{1}$ are the integration constant (rest mass of the
fermionic particle) and the perturbed mass$,$ respectively. Furthermore, $b$
is the dimensionless real constant to be set to zero for the constant mass
case (\textit{i.e.}, $m_{1}=0$) and $\lambda _{0}$ is the compton-like
wavelength in $fm$ units. It is worth mentioning that the above choice of
the mass function of Coulombic form [48,49] is mostly suitable for modeling
the well-known pseudo-Coulomb (Kratzer-type) potential [54-56]. The
interaction field has much impact on the choice of the mass function which,
in the present case, is inveresely proportional to the distance between the
two nuclei at short distances $m(r)\sim \frac{1}{r}$ and constant at long
distances $m(r\rightarrow \infty )\simeq m_{0}$.

\subsection{Spin symmetric solution of the Coulomb potential}

In the case of exact spin symmetry $S(r)\sim V(r)$ ($d\Delta (r)/dr=0,$
\textit{i.e}., $\Delta (r)=A=$ constant), eq. (20a) can be approximately
written as
\end{subequations}
\begin{equation}
\left\{ \frac{d^{2}}{dr^{2}}-\frac{\kappa \left( \kappa +1\right) }{r^{2}}-%
\frac{1}{\hbar ^{2}c^{2}}\left[ m(r)c^{2}+E_{n\kappa }-A\right] \left[
m(r)c^{2}-E_{n\kappa }+\Sigma (r)\right] \right\} F_{n\kappa }(r)=0,
\end{equation}%
where $\kappa =l$ for $\kappa <0$ and $\kappa =-\left( l+1\right) $ for $%
\kappa >0.$ The spin symmetric energy eigenvalues depend on $n$ and $\kappa
, $ \textit{i.e.}, $E_{n\kappa }=E(n,\kappa \left( \kappa +1\right) ).$ In
the last equation, the choice of $\Sigma (r)=2V(r)\rightarrow V(r)$ as
mentioned in ref. [12] enables one to reduce the resulting relativistic
solutions into their non-relativistic limit under appropriate
transformations. However, if we set $\Sigma (r)=2V(r),$ this allows us to
compare the resulting Dirac's solutions with those ones derived before in
refs. [47-49] regarding the solution of the KG equation for equal mixings $%
S(r)=V(r)$ (i.e., $\Delta =0$)$.$

Substituting eqs. (21) and (22) into eq. (23), allows us to decompose the
Dirac equation and then leading to obtain the Schr\"{o}dinger-like equation
in the spherical coordinates for the upper-spinor component $F_{n\kappa
}(r), $
\begin{equation}
\left[ \frac{d^{2}}{dr^{2}}+\left( \frac{-\varepsilon _{n\kappa
}^{2}r^{2}+\beta r-\gamma }{r^{2}}\right) \right] F_{n\kappa }(r)=0,
\end{equation}%
where we have defined
\begin{subequations}
\begin{equation}
\varepsilon _{n\kappa }=\frac{1}{\hbar c}\sqrt{m_{0}^{2}c^{4}-E_{n\kappa
}^{2}-A\left( m_{0}c^{2}-E_{n\kappa }\right) }>0,
\end{equation}%
\begin{equation}
\beta =\frac{1}{\hbar c}\left[ 2q\left( m_{0}c^{2}+E_{n\kappa }-A\right)
+b\left( A-2m_{0}c^{2}\right) \right] ,\text{ }
\end{equation}%
\begin{equation}
\gamma =b\left( b-2q\right) +\kappa \left( \kappa +1\right) .
\end{equation}%
For instance, when $A=0,$ the following constraint $E_{nl}<m_{0}c^{2}$ must
be fulfilled for bound state solutions. The quantum condition is obtained
from the finiteness of the solution at infinity and at the origin point
(i.e., $F_{n\kappa }(0)=F_{n\kappa }(\infty )=0$)$.$ In order to solve eq.
(24) by means of the NU method, we should compare it with eq. (1). The
following values for the parameters are found as
\end{subequations}
\begin{equation}
\widetilde{\tau }(r)=0,\text{ }\ \sigma (r)=r,\text{ }\ \widetilde{\sigma }%
(r)=-\varepsilon _{n\kappa }^{2}r^{2}+\beta r-\gamma .
\end{equation}%
Further, inserting these values into eq. (10), we obtain
\begin{equation}
\pi (r)=\frac{1}{2}\pm \frac{1}{2}\sqrt{1+4\gamma +4(k-\beta )r+4\varepsilon
_{n\kappa }^{2}r^{2}}.
\end{equation}%
The discriminant of the square root must be set equal to zero, \textit{i.e}%
., $1+4\gamma +4(k-\beta )r+4\varepsilon _{n\kappa }^{2}r^{2}=0.$ Hence, we
find the following two constants:%
\begin{equation}
k_{1,2}=\beta \pm \varepsilon _{n\kappa }\sqrt{1+4\gamma }.
\end{equation}%
In this regard, we can find the possible functions for $\pi (r)$ as
\begin{equation}
\pi (r)=\left\{
\begin{array}{cc}
\frac{1}{2}\pm \left[ \varepsilon _{n\kappa }r+\frac{1}{2}\sqrt{1+4\gamma }%
\right] & \text{\ for }k_{1}=\beta +\varepsilon _{n\kappa }\sqrt{1+4\gamma },
\\
\frac{1}{2}\pm \left[ \varepsilon _{n\kappa }r-\frac{1}{2}\sqrt{1+4\gamma }%
\right] & \text{\ for }k_{2}=\beta -\varepsilon _{n\kappa }\sqrt{1+4\gamma }.%
\end{array}%
\right.
\end{equation}%
According to the NU method, one of the four values of the polynomial $\pi
(r) $ is just proper to obtain the bound states because $\tau (r)$ has a
negative derivative. Therefore, the selected forms of $\pi (r)$ and $k$ take
the following particular values%
\begin{equation}
\pi (r)=\frac{1}{2}\left( 1+\sqrt{1+4\gamma }\right) -\varepsilon _{n\kappa
}r,\text{ }k=\beta -\varepsilon _{n\kappa }\sqrt{1+4\gamma },
\end{equation}%
to obtain%
\begin{equation}
\tau (r)=1+\sqrt{1+4\gamma }-2\varepsilon _{n\kappa }r,\text{ }\tau ^{\prime
}(r)=-2\varepsilon _{n\kappa }<0,
\end{equation}%
where prime denotes the derivative with respect to $r.$ In addition, after
using eqs. (6) and (9) together with the assignments given in eqs. (26),
(30) and (31), the following expressions for $\lambda $ are obtained as%
\begin{equation}
\lambda _{n}=\lambda =2n\varepsilon _{n\kappa },\text{ }n=0,1,2,\cdots ,
\end{equation}%
\begin{equation}
\lambda =\beta -\varepsilon _{n\kappa }\left( 1+\sqrt{1+4\gamma }\right) .
\end{equation}%
Now, taking $\lambda _{n}=\lambda ,$ we can solve the above equations to
obtain the energy equation for the Coulomb-like potential with spin-symmetry
in the Dirac theory,%
\begin{equation}
m_{0}^{2}c^{4}-E_{n\kappa }^{2}-A\left( m_{0}c^{2}-E_{n\kappa }\right) =
\left[ \frac{q\left( m_{0}c^{2}+E_{n\kappa }-A\right) +b\left( \frac{A}{2}%
-m_{0}c^{2}\right) }{n+\delta +1}\right] ^{2},
\end{equation}%
where
\begin{equation}
\delta =\sqrt{\left( \frac{1}{2}+\kappa \right) ^{2}+b(b-2q)}-\frac{1}{2}.
\end{equation}%
For spatially-dependent mass case, \textit{i.e.}, $b\neq 0,$ the bound state
solutions of the system are determined by means of the parameters $q$ and $%
b. $ It is not difficult to conclude that all bound-states appear in pairs,
two energy solutions are valid for the particle $E^{p}=E_{n\kappa }^{+}$ and
the second one corresponds to the anti-particle energy $E^{a}=E_{n\kappa
}^{-}$ in the Coulomb-like field.

Let us now find the corresponding eigenfunctions for this system. Using eqs.
(3) and (8), we find%
\begin{equation}
\rho (r)=r^{2\delta +1}e^{-2\varepsilon _{n\kappa }r},
\end{equation}%
\begin{equation}
\phi (r)=r^{\delta +1}e^{-\varepsilon _{n\kappa }r},
\end{equation}%
and further substituting eqs. (26) and (36) into eq. (7), we get%
\begin{equation}
y_{n}(r)=A_{n}r^{-(2\delta +1)}e^{2\varepsilon _{n\kappa }r}\frac{d^{n}}{%
dr^{n}}\left[ r^{2\delta +n+1}e^{-2\varepsilon _{n\kappa }r}\right] \sim
L_{n}^{2\delta +1}(2\varepsilon _{n\kappa }r),
\end{equation}%
where $L_{n}^{\alpha }(x)$ is the generalized Laguerre polynomials. The
relation $F_{n\kappa }(r)=\phi (r)y_{n}(r)$ gives the radial upper spinor
wave function:%
\begin{equation}
F_{n\kappa }(r)=\mathcal{N}r^{\delta +1}e^{-\varepsilon _{n\kappa
}r}L_{n}^{2\delta +1}(2\varepsilon _{n\kappa }r).
\end{equation}%
The above upper-spinor component satisfies the restriction condition for the
bound states, \textit{i.e.}, $\delta >0$ and $\varepsilon _{n\kappa }>0.$
Using the normalization condition $\int_{0}^{\infty }u(r)^{2}dr=1$ and the
orthogonality relation of the generalized Laguerre polynomials $%
\int_{0}^{\infty }x^{\alpha +1}e^{-x}\left[ L_{n}^{(\alpha )}(x)\right]
^{2}dx=\left( 2n+\alpha +1\right) \frac{\Gamma (n+\alpha +1)}{n!}$, the
normalizing factor $\mathcal{N}$ can be found as [57-59]%
\begin{equation}
\mathcal{N}=\sqrt{\frac{n!\left( 2\varepsilon _{n\kappa }\right) ^{2\delta
+3}}{2(n+\delta +1)\Gamma (n+2\delta +2)}},
\end{equation}%
where $\varepsilon _{n\kappa }$ and $\delta $ are given in eqs. (25a) and
(35), respectively.

On the other hand, it is worth noting that if we take the particular values
of the coefficients displayed in Table 1, then the expressions given in
Appendix A can be calculated for eq. (24) and tested with the ones
calculated before in a very simple way.

Let us now study some special cases of much concern to the reader. The
following cases are found to be identical with the solution of the
spatially-dependent mass and constant mass KG equation for the case spin
symmetry case $S(r)=V(r)$ ($\Delta (r)=\Sigma (r)=0$) [29,51]$.$

(1) If we consider the case when $S(r)=V(r),$ \textit{i.e.}, $q_{s}=q_{v}=q$
or $A=0,$ then eq. (34) can be reduced to the following forms:
\begin{subequations}
\begin{equation}
E_{n\kappa }^{p}=\frac{\left[ q\left( b-q\right) +B_{n\kappa }\sqrt{%
B_{n\kappa }^{2}-b\left( b-2q\right) }\right] }{q^{2}+B_{n\kappa }^{2}}%
m_{0}c^{2},
\end{equation}%
and%
\begin{equation}
\text{ }E_{n\kappa }^{a}=\frac{\left[ q\left( b-q\right) -B_{nl}\sqrt{%
B_{n\kappa }^{2}-b\left( b-2q\right) }\right] }{q^{2}+B_{n\kappa }^{2}}%
m_{0}c^{2},
\end{equation}%
where $B_{n\kappa }$ is defined by
\end{subequations}
\begin{equation}
B_{n\kappa }=\left( n+\frac{1}{2}+\sqrt{\left( \frac{1}{2}+\kappa \right)
^{2}+b(b-2q)}\right) .
\end{equation}%
\ Obviously, the bound-state solutions of the particle and anti-particle do
exist.

(2) When the mass is constant, i.e., $b=0$ ($m_{1}=0$)$,$ we have%
\begin{equation}
E_{n\kappa }^{p}=\frac{\left( n+\kappa +1\right) ^{2}-q^{2}}{\left( n+\kappa
+1\right) ^{2}+q^{2}}m_{0}c^{2}\text{ and \ }E_{n\kappa }^{a}=-m_{0}c^{2},%
\text{ }n=0,1,2,\cdots ,\text{ }\kappa =\pm 1,\pm 2,\cdots ,
\end{equation}%
which gives%
\begin{equation}
E_{n\kappa }^{p}=\left\{
\begin{array}{cc}
\frac{\left( n-l\right) ^{2}-q^{2}}{\left( n-l\right) ^{2}+q^{2}}m_{0}c^{2},
& \kappa >0 \\
\frac{\left( n+l+1\right) ^{2}-q^{2}}{\left( n+l+1\right) ^{2}+q^{2}}%
m_{0}c^{2}, & \kappa <0%
\end{array}%
\right. \text{ and }E_{n\kappa }^{a}=-m_{0}c^{2},\text{ }l=0,1,2,\cdots ,
\end{equation}%
where $n$ and $\kappa $ signify the usual radial and spin-orbit quantum
numbers. The particle has bound state solution whereas anti-particle has
continuum solution for all states. Further, the wave functions for particle
turn to be
\begin{equation*}
F_{n\kappa }^{p}(r)=\sqrt{\frac{n!\left( \frac{2}{\hbar c}\sqrt{%
m_{0}^{2}c^{4}-E_{n\kappa }^{p2}}\right) ^{2\kappa +3}}{2(n+\kappa +1)\Gamma
(n+2\kappa +2)}}r^{\kappa +1}\exp \left( -\frac{2}{\hbar c}\sqrt{%
m_{0}^{2}c^{4}-E_{n\kappa }^{p2}}r\right)
\end{equation*}%
\begin{equation}
\times L_{n}^{2\kappa +1}\left( \frac{2}{\hbar c}\sqrt{m_{0}^{2}c^{4}-E_{n%
\kappa }^{p2}}r\right) ,\text{ }\kappa >0,\text{ }E_{n\kappa
}^{p}<m_{0}c^{2},
\end{equation}%
whereas the anti-particle has no wave functions.

(3) When the potential coupling constant is set to $q=b/2,$ the spectrum of
the spatially-dependent mass Dirac particle in potential field $%
q_{s}=q_{v}=q $ is similar to the spectrum of constant mass Dirac particle
in the potential fields $q_{s}=-q_{v},$ that is,

\begin{subequations}
\begin{equation}
E_{n\kappa }^{p}=m_{0}c^{2},
\end{equation}%
\begin{equation}
\text{ }E_{n\kappa }^{a}=-\frac{\left( n+\kappa +1\right) ^{2}-q^{2}}{\left(
n+\kappa +1\right) ^{2}+q^{2}}m_{0}c^{2}.
\end{equation}

(4) When the potential coupling constant is set to $q=b/2,$ the spectrum of
the spatially-dependent mass Dirac particle in potential field $q_{s}=-q_{v}$
is similar to the spectrum of constant mass Dirac particle in the potential
fields $q_{s}=q_{v}=q,$ that is,
\end{subequations}
\begin{subequations}
\begin{equation}
E_{nl}^{p}=\frac{\left( n+\kappa +1\right) ^{2}-q^{2}}{\left( n+\kappa
+1\right) ^{2}+q^{2}}m_{0}c^{2},
\end{equation}%
\begin{equation}
E_{nl}^{a}\text{ }=-m_{0}c^{2}.
\end{equation}

\subsection{Pseudospin symmetric solution of the Coulomb potential}

In the presence of the pseudospin symmetry $S(r)\sim -V(r)$ (\textit{i.e}., $%
d\Sigma (r)/dr=0,$ or $\Sigma (r)=A=$ constant), eq. (20b) can be exactly
written as
\end{subequations}
\begin{equation}
\left\{ \frac{d^{2}}{dr^{2}}-\frac{\kappa \left( \kappa -1\right) }{r^{2}}-%
\frac{1}{\hbar ^{2}c^{2}}\left[ m(r)c^{2}+E_{n\kappa }-\Delta (r)\right] %
\left[ m(r)c^{2}-E_{n\kappa }+A\right] \right\} G_{n\kappa }(r)=0,
\end{equation}%
where the energy eigenvalues $E_{n\kappa }$ depend only on $n$ and $\kappa ,$
\textit{i.e.}, $E_{n\kappa }=E(n,\kappa (\kappa -1)).$ We take the
difference of the potential as $2V(r)$ which allows us to compare our final
results with the ones calculated for the KG equation derived in ref. [49].
Substituting%
\begin{equation*}
\Delta (r)=V(r)=-\frac{\hbar cq_{v}}{r},q_{v}=-q,r\neq 0,
\end{equation*}%
together with eq. (22) into the last equation, we obtain a Schr\"{o}dinger
euation for the lower component $G_{n\kappa }(r),$%
\begin{equation}
\left[ \frac{d^{2}}{dr^{2}}+\left( \frac{-\widetilde{\varepsilon }_{n\kappa
}^{2}r^{2}+\widetilde{\beta }r-\widetilde{\gamma }}{r^{2}}\right) \right]
G_{n\kappa }(r)=0,\text{ }G_{n\kappa }(0)=G_{n\kappa }(\infty )=0,
\end{equation}%
where we have defined
\begin{subequations}
\begin{equation}
\widetilde{\varepsilon }_{n\kappa }=\frac{1}{\hbar c}\sqrt{%
m_{0}^{2}c^{4}-E_{n\kappa }^{2}+A\left( m_{0}c^{2}+E_{n\kappa }\right) }>0,
\end{equation}%
\begin{equation}
\widetilde{\beta }=-\frac{1}{\hbar c}\left[ 2q\left( m_{0}c^{2}-E_{n\kappa
}+A\right) +b\left( 2m_{0}c^{2}+A\right) \right] ,\text{ }
\end{equation}%
\begin{equation}
\widetilde{\gamma }=b\left( b+2q\right) +\kappa \left( \kappa +1\right) .%
\text{ }
\end{equation}%
To avoid repetition in the solution of eq. (49), a first inspection for the
relationship between the present set of parameters $(\widetilde{\varepsilon }%
_{n\kappa },\widetilde{\beta },\widetilde{\gamma })$ and the previous set $%
(\varepsilon _{n\kappa },\beta ,\gamma )$ tells us that the negative energy
solution for pseudospin symmetry, where $S(r)=-V(r),$ can be obtained
directly from those of the positive energy solution above for spin symmetry
using the parameter map [50-52]:
\end{subequations}
\begin{equation}
F_{n\kappa }(r)\leftrightarrow G_{n\kappa }(r),V(r)\rightarrow -V(r)\text{
(or }q\rightarrow -q\text{)},\text{ }E_{n\kappa }\rightarrow -E_{n\kappa }%
\text{ and }A\rightarrow -A.
\end{equation}%
Following the previous results with the above transformations, we finally
arrive at the energy equation,%
\begin{equation}
m_{0}^{2}c^{4}-E_{n\kappa }^{2}+A\left( m_{0}c^{2}+E_{n\kappa }\right) =
\left[ \frac{q\left( m_{0}c^{2}-E_{n\kappa }+A\right) +b\left( m_{0}c^{2}+%
\frac{A}{2}\right) }{n+\frac{1}{2}+\sqrt{\left( \kappa -\frac{1}{2}\right)
^{2}+b(b+2q)}}\right] ^{2},
\end{equation}%
and lower-spinor component wave function,%
\begin{equation}
G_{n\kappa }(r)=\sqrt{\frac{n!\left( 2\widetilde{\varepsilon }_{n\kappa
}\right) ^{2\eta +3}}{2(n+\eta +1)\Gamma (n+2\eta +2)}}r^{\eta +1}e^{-%
\widetilde{\varepsilon }_{n\kappa }r}L_{n}^{2\eta +1}(2\widetilde{%
\varepsilon }_{n\kappa }r),
\end{equation}%
where $\eta =\sqrt{\left( \kappa -\frac{1}{2}\right) ^{2}+b(b+2q)}-\frac{1}{2%
}.$ The finiteness of the above wave functions for the bound states is
achieved once $\eta >0$ and $\widetilde{\varepsilon }_{n\kappa }>0.$ Let us
investigate the following specific cases of interest.

(1) Considering the case when $S(r)=-V(r),$ \textit{i.e.}, $q_{s}=-q_{v}=q$
or $A=0,$ then eq. (52) can be reduced to the following forms:
\begin{subequations}
\begin{equation}
E_{n\kappa }^{p}=\frac{\left[ -q\left( b+q\right) +B_{n\kappa }\sqrt{%
B_{n\kappa }^{2}-b\left( b+2q\right) }\right] }{q^{2}+B_{n\kappa }^{2}}%
m_{0}c^{2},
\end{equation}%
\begin{equation}
\text{ }E_{n\kappa }^{a}=\frac{\left[ -q\left( b+q\right) -B_{nl}\sqrt{%
B_{n\kappa }^{2}-b\left( b+2q\right) }\right] }{q^{2}+B_{n\kappa }^{2}}%
m_{0}c^{2},
\end{equation}%
where $B_{n\kappa }$ is defined by
\end{subequations}
\begin{equation}
B_{n\kappa }=\left( n+\frac{1}{2}+\sqrt{\left( \kappa -\frac{1}{2}\right)
^{2}+b(b+2q)}\right) .
\end{equation}%
\ We note that the bound state solutions of the particle and anti-particle
are available.

(2) When the mass is constant, i.e., $b=0,$ then we obtain%
\begin{equation}
E_{n\kappa }^{p}=\frac{\left( n+\kappa \right) ^{2}-q^{2}}{\left( n+\kappa
\right) ^{2}+q^{2}}m_{0}c^{2},\text{ }E_{n\kappa }^{a}=-m_{0}c^{2},\text{ }%
n=0,1,2,\cdots ,\text{ }\kappa =\pm 1,\pm 2,\cdots ,
\end{equation}%
and the lower-spinor wave functions:%
\begin{equation*}
G_{n\kappa }(r)=\sqrt{\frac{n!\left( \frac{2}{\hbar c}\sqrt{%
m_{0}^{2}c^{4}-E_{n\kappa }^{2}}\right) ^{2\kappa +1}}{2(n+\kappa )\Gamma
(n+2\kappa )}}r^{\kappa }e^{-\frac{1}{\hbar c}\sqrt{m_{0}^{2}c^{4}-E_{n%
\kappa }^{2}}r}
\end{equation*}%
\begin{equation}
\times L_{n}^{2\kappa -1}\left( \frac{2}{\hbar c}\sqrt{m_{0}^{2}c^{4}-E_{n%
\kappa }^{2}}r\right) .
\end{equation}%
where $\kappa =1,2,3,\cdots ,$ and $E_{n\kappa }<m_{0}c^{2}.$

\section{Discussions}

In this section, we are going to study two special cases of the energy
eigenvalues given by eqs. (34) and (52). First, let us study $s$-wave case $%
l=0$ ($\kappa =-1$) and $\widetilde{l}=0$ ($\kappa =1$) case%
\begin{equation}
m_{0}^{2}c^{4}-E_{n,-1}^{2}-A\left( m_{0}c^{2}-E_{n,-1}\right) =\left[ \frac{%
2q\left( m_{0}c^{2}+E_{n,-1}-A\right) +b\left( A-2m_{0}c^{2}\right) }{2n+1+%
\sqrt{1+4b(b-2q)}}\right] ^{2},
\end{equation}%
\begin{equation}
m_{0}^{2}c^{4}-E_{n,1}^{2}+A\left( m_{0}c^{2}+E_{n,1}\right) =\left[ \frac{%
2q\left( m_{0}c^{2}-E_{n,1}+A\right) +b\left( A+2m_{0}c^{2}\right) }{2n+1+%
\sqrt{1+4b(b+2q)}}\right] ^{2}.
\end{equation}%
For constant mass case, if one set $A=0$ and $b=0$ into eqs. (58) and (59),
we obtain for spin and pseudo-spin symmetric Dirac theory,
\begin{subequations}
\begin{equation}
E_{n,-1}=\frac{\left( n+1\right) ^{2}-q^{2}}{\left( n+1\right) ^{2}+q^{2}}%
m_{0}c^{2},
\end{equation}%
\end{subequations}
\begin{subequations}
\begin{equation}
E_{n,1}=-\frac{\left( n+1\right) ^{2}-q^{2}}{\left( n+1\right) ^{2}+q^{2}}%
m_{0}c^{2}.
\end{equation}%
respectively, which are identical with the $s$-wave results in ref. [48] for
$S(r)=V(r)$ and $S(r)=-V(r),$ respectively. Let us now discuss the
non-relativistic limit of the energy eigenvalues and wave functions of our
solution for the spatially dependent mass and for the constant mass Schr\"{o}%
dinger equation. If we take $A=0$ and put $2q\rightarrow q$ [12]$,$ \textit{%
i.e}., $S(r)=V(r)=\Sigma (r),$ the non-relativistic limits of energy
equation (34) and wave functions (39) under the following transformations $%
E_{n\kappa }+m_{0}c^{2}\approx 2\mu /\hbar ^{2}$ and $E_{n\kappa
}-m_{0}c^{2}\approx E_{nl}$ [27,50] become
\end{subequations}
\begin{equation}
E_{nl}=-\frac{\mu }{2\hbar ^{2}}\frac{(q-b)^{2}}{\left( n+\frac{1}{2}+\sqrt{%
\left( \frac{1}{2}+l\right) ^{2}+b(b-2q)}\right) ^{2}},
\end{equation}%
\begin{equation}
F_{nl}(r)=\mathcal{N}r^{\frac{1}{2}\left( 1+\sqrt{\left( 2l+1\right)
^{2}+4b(b-q)}\right) }e^{-\sqrt{-\frac{2\mu }{\hbar ^{2}}E_{nl}}r}L_{n}^{%
\sqrt{\left( 2l+1\right) ^{2}+4b(b-q)}}\left( 2\sqrt{-\frac{2\mu }{\hbar ^{2}%
}E_{nl}}r\right) ,
\end{equation}%
respectively, where%
\begin{equation}
\mathcal{N}=\sqrt{\frac{n!\left( 2\sqrt{-\frac{2\mu }{\hbar ^{2}}E_{nl}}%
\right) ^{\sqrt{\left( 2l+1\right) ^{2}+4b(b-q)}+2}}{\left( 2n+1+\sqrt{%
\left( 2l+1\right) ^{2}+4b(b-q)}\right) \Gamma \left( n+1+\sqrt{\left(
2l+1\right) ^{2}+4b(b-q)}\right) }}.
\end{equation}%
Furthermore, when mass becomes constant $(b=0),$ we find the well-known
non-relativistic energy spectrum and wave functions of the Coulomb problem,%
\begin{equation}
E_{nl}=-\frac{\mu }{2\hbar ^{2}}\frac{q^{2}}{\left( n+l+1\right) ^{2}},\text{
}q=Z\alpha
\end{equation}%
\begin{equation}
F_{nl}(r)=\mathcal{N}r^{l+1}e^{-\sqrt{-\frac{2\mu }{\hbar ^{2}}E_{nl}}%
r}L_{n}^{\left( 2l+1\right) }\left( 2\sqrt{-\frac{2\mu }{\hbar ^{2}}E_{nl}}%
r\right) ,\text{ }\mathcal{N}=\sqrt{\frac{n!\left( 2\sqrt{-\frac{2\mu }{%
\hbar ^{2}}E_{nl}}\right) ^{2l+3}}{2\left( n+l+1\right) \Gamma \left(
n+2l+2\right) }}
\end{equation}%
which are identical to those given in refs. [60-62].

\section{Conclusions}

To summarize, we have presented the bound state solutions of the
spatially-dependent mass Dirac equation with the Coulomb-like field
potential under the conditions of the spin symmetry and pseudospin symmetry.
We have obtained an explicit expressions for energy eigenvalues and
associated wave functions for arbitrary spin-orbit $\kappa $ state with a
specific choice for the mass function that provides an exact solution to the
pseudospin symmetric Dirac equation and an approximate solution for the spin
symmetric Dirac equation. This suitable choice of mass function which is in
the form of Coulomb-like potential enables us to solve the
spatially-dependent Dirac equation analytically and reduces it to the
constant mass solution as well as when $b=0$ ($m_{1}=0$). The most stringent
interesting result is that the present spin and pseudo-spin symmetries can
be easily reduced to the previously found Klein-Gordon solution once $%
S(r)=V(r)$ and $S(r)=-V(r)$ (\textit{i.e}., when $A=0$)$,$ respectively. The
resulting solutions of the wave functions are being expressed in terms of
the generalized Laguerre polynomials. Obviously, when the coupling potential
parameters are adjusted to some specific values, particularily when $q=b/2,$
the spectra of the mass varying Dirac particle for the case $q_{s}=q_{v}$ ($%
q_{s}=-q_{v}$) turn to become similar to the spectra of the constant mass
Dirac particle for the case $q_{s}=-q_{v}$ ($q_{s}=q_{v}$), respectively. In
the limit of constant mass ($b=0$), the solution for the energy eigenvalues
and wave functions are reduced to those ones given in literature. Also, when
spin-orbit quantum number $\kappa =0,$ the problem reduces to $s$-waves
solution as in [48].

\acknowledgments We highly appreciate the partial support provided by the
Scientific and Technological Research Council of Turkey (T\"{U}B\.{I}TAK).

\newpage \appendix

\section{Parametric Generalization of the NU Method}

Our systematical derivations hold for any potential form.

(i) The analytic results for the key polynomials [47]:
\begin{equation}
\pi (r)=c_{5}+\sqrt{c_{9}}-\frac{1}{c_{3}}\left( c_{4}\sqrt{c_{9}}+\sqrt{%
c_{10}}-c_{3}c_{6}\right) r,
\end{equation}%
\begin{equation}
k=-\frac{1}{c_{3}^{2}}\left( c_{3}c_{8}+2c_{4}c_{9}+2\sqrt{c_{9}c_{10}}%
\right) ,
\end{equation}%
\begin{equation}
\tau (r)=c_{3}+2\sqrt{c_{9}}-\frac{2}{c_{3}}\left( c_{3}c_{4}+c_{4}\sqrt{%
c_{9}}+\sqrt{c_{10}}\right) r,
\end{equation}%
\begin{equation}
\tau ^{\prime }(r)=-\frac{2}{c_{3}}\left( c_{3}c_{4}+c_{4}\sqrt{c_{9}}+\sqrt{%
c_{10}}\right) <0.
\end{equation}%
(ii) The energy equation:%
\begin{equation*}
c_{2}n-\left( 2n+1\right) c_{6}+\frac{1}{c_{3}}\left( 2n+1\right) \left(
\sqrt{c_{10}}+c_{4}\sqrt{c_{9}}\right)
\end{equation*}%
\begin{equation}
+n\left( n-1\right) c_{4}+\frac{1}{c_{3}^{2}}\left( c_{3}c_{8}+2c_{4}c_{9}+2%
\sqrt{c_{9}c_{10}}\right) =0.
\end{equation}%
(iii) The wave functions:%
\begin{equation}
\rho (r)=r^{c_{11}}(c_{3}-c_{4}r)^{c_{12}},
\end{equation}%
\begin{equation}
\phi (r)=r^{c_{13}}(c_{3}-c_{4}r)^{c_{14}},
\end{equation}%
\begin{equation}
y_{n}(r)=P_{n}^{\left( c_{11},c_{12}\right) }(c_{3}-c_{4}r),
\end{equation}%
\begin{equation}
\psi _{n}(r)=\phi (r)y_{n}(r)=\mathcal{A}%
_{n}r^{c_{13}}(c_{3}-c_{4}r)^{c_{14}}P_{n}^{\left( c_{11},c_{12}\right)
}(c_{3}-2c_{4}r),
\end{equation}%
where $P_{n}^{\left( a,b\right) }(c_{3}-c_{4}r)$ are the Jacobi polynomials
and $\mathcal{A}_{n}$ is a normalizing factor.

When $c_{4}=0,$ the Jacobi polynomial turn to be the generalized Laguerre
polynomial and the constants relevant to this polynomial change are%
\begin{equation}
\lim_{c_{4}\rightarrow
0}P_{n}^{(c_{11},c_{12})}(c_{3}-2c_{4}r)=L_{n}^{c_{11}}(c_{15}r),
\end{equation}%
\begin{equation}
\lim_{c_{4}\rightarrow 0}(c_{3}-c_{4}r)^{c_{14}}=e^{-c_{16}r},
\end{equation}%
where $L_{n}^{c_{11}}(c_{15}r)$ are the generalized Laguerre polynomials and
$\mathcal{B}_{n}$ is a normalizing factor.

(iv) The relevant coefficients $c_{i}$ ($i=5,6,\cdots ,16$) are given as
follows:%
\begin{equation}
c_{5}=\frac{1}{2}\left( c_{3}-c_{1}\right) ,\text{ }c_{6}=\frac{1}{2}\left(
c_{2}-2c_{4}\right) ,\text{ }c_{7}=c_{6}^{2}+\xi _{1},
\end{equation}%
\begin{equation*}
\text{ }c_{8}=2c_{5}c_{6}-\xi _{2},\text{ }c_{9}=c_{5}^{2}+\xi _{3},
\end{equation*}%
\begin{equation}
c_{10}=c_{4}\left( c_{3}c_{8}+c_{4}c_{9}\right) +c_{3}^{2}c_{7},
\end{equation}%
\begin{equation}
c_{11}=\frac{2}{c_{3}}\sqrt{c_{9}},\text{ }c_{12}=\frac{2}{c_{3}c_{4}}\sqrt{%
c_{10}},
\end{equation}%
\begin{equation}
c_{13}=\frac{1}{c_{3}}\left( c_{5}+\sqrt{c_{9}}\right) ,\text{ }c_{14}=\frac{%
1}{c_{3}c_{4}}\left( \sqrt{c_{10}}-c_{4}c_{5}-c_{3}c_{6}\right) ,
\end{equation}%
\begin{equation}
c_{15}=\frac{2}{c_{3}}\sqrt{c_{10}},\text{ }c_{16}=\frac{c_{15}}{2}.
\end{equation}%
\newpage\

{\normalsize 
}

\bigskip

\bigskip

\baselineskip= 2\baselineskip
\bigskip \newpage

\bigskip

\bigskip {\normalsize 
}

\baselineskip= 2\baselineskip

\bigskip
\begin{table}[tbp]
\caption{The specific values for the parametric constants necessary for
calculating the energy eigenvalues and eigenfunctions of the spin symmetry
Dirac wave equation (23).}%
\begin{tabular}{llll}
\tableline Constant & Analytic value & Constant & Analytic value \\
\tableline$c_{1}$ & $0$ & $c_{2}$ & $0$ \\
$c_{3}$ & $1$ & c$_{4}$ & $0$ \\
$c_{5}$ & $\frac{1}{2}$ & $c_{6}$ & $0$ \\
$c_{7}$ & $\varepsilon _{n\kappa }^{2}$ & $c_{8}$ & $-\beta $ \\
$c_{9}$ & $\frac{1}{4}+\gamma $ & $c_{10}$ & $\varepsilon _{n\kappa }^{2}$
\\
$c_{11}$ & $1+2\delta $ & $c_{12}=c_{15}$ & 2$\varepsilon _{n\kappa }$ \\
$c_{13}$ & $1+\delta $ & $c_{14}=c_{16}$ & $\varepsilon _{n\kappa }$ \\
$\xi _{1}$ & $\varepsilon _{n\kappa }^{2}$ & $\xi _{2}$ & $\beta $ \\
$\xi _{3}$ & $\gamma $ &  &  \\
\tableline &  &  &
\end{tabular}%
\end{table}
\

\end{document}